\documentclass[letters,fleqn,usenatbib]{mnras}
\usepackage[T1]{fontenc}
\usepackage{ae,aecompl}
\usepackage{graphicx}	
\usepackage{amsmath}	
\usepackage{amssymb}	
\usepackage{booktabs}

\newcommand{\be}{\begin{equation}}
\newcommand{\ee}{\end{equation}}
\newcommand{\bq}{\begin{eqnarray}}
\newcommand{\eq}{\end{eqnarray}}

\title[Fundamental cosmology from ANDES]{Fundamental cosmology from ANDES precision spectroscopy}

\author[C. M. J. Marques {\it et al.}]{
C. M. J. Marques,$^{1,2,3,4}$\thanks{E-mail: Catarina.Marques@astro.up.pt (CMJM)}
C. J. A. P. Martins$^{1,3}$\thanks{E-mail: Carlos.Martins@astro.up.pt (CJAPM)}
and C. S. Alves$^{1,5}$\thanks{E-mail: catarina.alves.18@ucl.ac.uk (CSA)}
\\
$^{1}$Centro de Astrof\'{\i}sica da Universidade do Porto, Rua das Estrelas, 4150-762 Porto, Portugal\\
$^{2}$Faculdade de Ci\^encias e Tecnologia, Universidade Nova de Lisboa, 2829-516 Caparica, Portugal\\
$^{3}$Instituto de Astrof\'{\i}sica e Ci\^encias do Espa\c co, CAUP, Rua das Estrelas, 4150-762 Porto, Portugal\\
$^{4}$Faculdade de Ci\^encias, Universidade do Porto, Rua do Campo Alegre, 4150-007 Porto, Portugal\\
$^{5}$Department of Physics and Astronomy, University College London, Gower Street, London WC1E 6BT, United Kingdom
}

\date{Accepted XXX. Received YYY; in original form ZZZ}

\pubyear{2022}

\begin{document}
\label{firstpage}
\pagerange{\pageref{firstpage}--\pageref{lastpage}}
\maketitle

\begin{abstract}
Fundamental cosmology observations, such as the detection of the redshift drift and tests of the universality of physical laws, are key science and design drivers of the ArmazoNes high Dispersion Echelle Spectrograph (ANDES), an Extremely Large Telescope instrument. While separate forecasts for each of them have been reported, we have developed Fisher Matrix based forecast tools combining both of these observables. We demonstrate the synergies between the two ANDES datasets, quantifying the improvements in cosmology and fundamental physics parameter constraints for two separate theoretical paradigms. We publicly release this forecast code, which is one of the tools for the optimisation of the ANDES observing strategy.
\end{abstract}

\begin{keywords}
Cosmology: cosmological parameters -- Cosmology: dark energy -- Cosmology: observations -- Methods: analytical -- Methods: statistical
\end{keywords}


\section{Introduction}

Precision spectroscopy has played a crucial role in XX century physics, e.g. in the development of quantum mechanics and the confirmation of quantum electrodynamics. In XXI century astrophysics, remarkable improvements in precision, accuracy and stability of high-resolution fiber-fed optical echelle spectrographs give them a unique role in fundamental physics tests and the search for new physics.

Absorption spectra of bright quasars are a powerful tool for astrophysical tests of fundamental physics, including the detection of the redshift drift \citep{Sandage,Liske} and tests of the universality of physical laws, through tests of the stability of the fine-structure constant, $\alpha=e^2/\hbar c$ \citep{Martins17}. In some cases, the same target yields both measurements, with the conceptual difference that the latter is a single-shot experiment while the former requires observations at two reasonably separate epochs. The two are also flagship science cases and design drivers for the high-resolution optical/infrared spectrograph, ANDES \citep{ANDES}, for the Extremely Large Telescope (ELT): a detection of the redshift drift requires sufficient precision and stability, while measurements of $\alpha$ require sufficient precision and accuracy \citep{HIRES}.

For the redshift drift, current upper limits are three orders of magnitude larger than the expected signal \citep{Darling,Cooke}; an ongoing ESPRESSO measurement will significantly improve them. There are several forecasts of the impact of ELT measurements, relying on the analysis of \citet{Liske}, e.g. \citet{Corasaniti,Balbi,Moraes,Martinelli,Alves,Esteves,Rocha}. Apart from the fundamental impact of seeing the universe expand in real time, the forecasts and related analyses presented in these earlier works show that although these cosmological constraints are not {\it per se} stronger than those of traditional observables, they play a key role in breaking parameter degeneracies (since they tend to be orthogonal to constraints from traditional probes), thereby improving constraints on the model's cosmological parameters. One aspect to bear in mind is that these previous works have studied the impact of combining redshift drift data with data from more traditional observables obtained by other instruments. Instead, the present work focuses on the combination of the redshift drift with a different observable obtained by the same instrument.

For the fine-structure constant many measurements exist, cf. the review by \citet{Martins17}, and the ESPRESSO spectrograph is enabling significant gains in precision and accuracy, which is now at the parts per million (ppm) level of relative variation \citep{Schmidt,Murphy}. Combining these measurements with cosmological data provides unique tests of fundamental physics, e.g. of Bekenstein models \citep{Bekenstein1,Bekenstein2}, and improves local tests of the Weak Equivalence Principle. They are expected to remain competitive with next-generation facilities \citep{AlvesAlpha,Boas,Euclid}.

Here we present an analysis of the \textit{joint} cosmological impact of redshift drift and $\alpha$ measurements, as expected to be provided by ANDES over a period of 10-20 years. We draw on previous literature and our ESPRESSO experience to generate simulated datasets for plausible ANDES specifications and study their cosmological impact, using Fisher Matrix Analysis techniques \citep{FMA1,FMA2}. We build upon codes previously developed by our team \citep{AlvesAlpha,Alves}; the present analysis uses a separate code\footnote{\href{https://github.com/CatarinaMMarques/FisherCosmology}{github.com/CatarinaMMarques/FisherCosmology}}, which was validated using the two earlier codes. We consider two classes of fiducial models: the aforementioned Bekenstein models, and also the Chevallier-Polarski-Linder (CPL) parametrization of dark energy \citep{CPL1,CPL2}. These are among the simplest parametric extensions of the $\Lambda$CDM model and are the two canonical representatives of the two broad classes of varying $\alpha$ models, further discussed below.


\section{Observables and Fiducial Models}

The redshift drift of an object which, in a time interval $\Delta t$, follows the cosmological expansion is \citep{Sandage}
\begin{equation}
    \frac{\Delta z}{\Delta t} = H_\mathrm{0}[1 + z - E(z)]\,,
\end{equation}
where for convenience we have defined $E(z)=H(z)/H_0$, although the true observable is the spectroscopic velocity
\begin{equation}
    \label{eq:spectr_velocity}
    \Delta v = \frac{c \Delta z}{1 + z} = c H_\mathrm{0} \Delta t \left[ 1 - \frac{E(z)}{1+z} \right]\,,
\end{equation}
since the measurement is effectively done in velocity space. The signal for the measurement comes mainly from the Lyman-$\alpha$ forest, but metal absorption lines redwards of it can also contribute  \citep{Liske}.

For our simulated ANDES redshift drift data, we follow \citet{Liske}, which finds that the spectroscopic velocity uncertainty can be approximated by
\begin{equation}
    \label{eq:uncLiske}
    \sigma_v = 1.35 \left(\frac{S/N}{2370}\right)^{-1} \left(\frac{N_\mathrm{Q}}{30}\right)^{-1/2} \left(\frac{1 + z_\mathrm{Q}}{5}\right)^{-\lambda}\mathrm{cm/s}, 
\end{equation}
where $\lambda = 1.7$ up to $z = 4$ and $\lambda = 0.9$ for $z > 4$,  $S/N$ is the spectral signal to noise ratio per 0.0125 \r{A} pixel, and $N_\mathrm{Q}$  is the number of objects observed at the effective redshift of the QSO, $z_\mathrm{Q}$. For consistency with other forecasts \citep{Alves,Esteves, Rocha}, we assume $N_{Q}=6$ quasars measured at each of five redshift bins $z=2.0, 2.5, 3.0, 3.5, 4.5$ (i.e., a total sample of 30 quasars) in a time span of $\Delta t=20$ years, with each spectrum having $S/N=3000$. While such a dataset is costly in telescope time, recent discoveries of bright high-redshift quasars can significantly reduce this time. We note that \citet{Boutsia} recently proposed a possible `Golden Sample' of 30 such quasars, but the redshift distribution of this new sample does not coincide with ours. We leave a trade-off study between possible sample choices for future work.

A dynamical scalar field is the natural alternative to the cosmological constant suggested by observations. This field is expected to be slow-rolling and will couple to the rest of the degrees of freedom unless an unknown symmetry is postulated to suppress the couplings. This leads to long-range forces and spacetime dependencies of nature's dimensionless couplings. In particular, a coupling to the electromagnetic sector implies a spacetime varying $\alpha$. Cosmological varying $\alpha$ models can be divided into two classes \citep{Martins17}.

In Class I models, we have a single new dynamical degree of freedom accounting for dark energy and a varying $\alpha$. It follows that the behaviour of $\alpha$ is parametrically determined. For a canonical scalar field with an equation of state $w_\mathrm{{\phi}} \geq -1$, the relative variation of $\alpha$ (as compared to its local laboratory value) is \citep{Calabrese}
\begin{equation}
    \label{eq:ClassImodels}
    \frac{\Delta \alpha}{\alpha}(z) = \zeta \int_{0}^{z} \sqrt{3\Omega_{\phi}(z')[1+w_{\phi}(z')]}\frac{dz'}{1+z'}\,,
\end{equation}
here $\zeta$ is the dimensionless coupling of the scalar field to the electromagnetic sector of the theory and $\Omega_{\phi} = \rho_{\phi} / (\rho_{\phi} + \rho_{m})$ is the fraction of the energy density of the scalar field relative to the
Universe's total energy density. Notice that if the coupling vanishes ($\zeta=0$) or if dark energy is a cosmological constant ($w_{\phi}=-1$) there will be no $\alpha$ variation. In other words, for these models the variation of $\alpha$ requires a non-trivial behaviour in both the gravitational and the particle physics sectors of the theory.

As fiducial model in this class we use the CPL parametrization \citep{CPL1,CPL2}, which postulates a dark energy equation of state
\begin{equation}
    w_\mathrm{{CPL}}(z) = w_0 + w_a \frac{z}{1 + z}\,,
\end{equation}
where $w_0$ is its present value and $w_a$ is the coefficient of the time-dependent term. With the further assumption of a flat universe (i.e., $\Omega_m + \Omega_\Lambda = 1$), the square of the rescaled Hubble parameter for this parametrization is
\begin{equation}
\label{eq:E2CPL}
    E^{2}(z) = \Omega_m (1+z)^{3} + (1- \Omega_m) (1+z)^{3(1+w_0+w_a)} \exp{\left( -\frac{3w_a z}{1+z} \right)}\,.
\end{equation}
Clearly, flat $\Lambda$CDM is recovered for $w_a = 0$ and $w_0 = -1$. With these assumptions the present drift rate of $\alpha$ is
\begin{equation}
\left(\frac{1}{H_0}\frac{\dot\alpha}{\alpha}\right)_0 = -\,\zeta \sqrt{3 (1-\Omega_m) (1+w_0)}\,,
\end{equation}
which again explicitly highlights the effects of the gravitational and fundamental physics sectors.

In Class II models, the field responsible for the varying $\alpha$ does not provide the dark energy, which in the simplest scenarios is assumed to be a cosmological constant. As fiducial model in this class, we use the Bekenstein model \citep[studied, e.g. in][]{AlvesAlpha,Bekenstein2} introduced by \citet{Olive}. Here different couplings to dark energy and matter sectors are allowed, and $\alpha$ evolves as
\begin{equation}
\begin{split}
    \label{eq:ClasseIImodels}
\frac{\Delta \alpha}{\alpha}(z) & = 2 \zeta_m \log(1+z) + \frac{2(\zeta_{\Lambda} - 2\zeta_m)}{3\sqrt{1-\Omega_m}} \left[ \log\left( \frac{1+\sqrt{1-\Omega_m}}{\sqrt{\Omega_m}}\right)\right. \\ & \left.- \sqrt{E^2(z)} \log\left( \frac{\sqrt{1-\Omega_m} + \sqrt{E^2(z)}}{\sqrt{\Omega_m (1+z)^3}} \right) \right]\,,
\end{split}
\end{equation}
where $\zeta_{\Lambda}$ and $\zeta_\mathrm{m}$ are the aforementioned dimensionless couplings. The rescaled Friedmann equation coincides with the $\Lambda$CDM one: it is Eq.  (\ref{eq:E2CPL}) with $w_a=0$ and $w_0=-1$, and again we recover $\Lambda$CDM when the two couplings vanish. In this case the present-day drift rate of $\alpha$ is
\be
\left(\frac{1}{H}\frac{\dot\alpha}{\alpha}\right)_0=-\zeta_\Lambda+2(\zeta_\Lambda-2\zeta_m)\frac{\Omega_m}{\sqrt{1-\Omega_m}}\log{\frac{1+\sqrt{1-\Omega_m}}{\sqrt{\Omega_m}}}\,.
\ee
In this class the gravitational sector is closer to the standard one, in the sense that the dark energy is due to a cosmological constant, but nevertheless the Einstein Equivalence Principle is violated. On the other hand, the particle physics sector is manifestly non-standard. Note also that the parameter dependencies in the current drift rate of $\alpha$ also differ from those in the Class I case above.

For simulated ANDES $\alpha$ measurements we take two different datasets, which we call Baseline and Optimistic. These measurements are obtained from metal absorption lines redwards of the Lyman-$\alpha$ forest. For the Baseline set we assume $15$ equally spaced measurements between $z=1.0$ and $z=2.5$, each with an uncertainty $\sigma_\alpha = 0.1$ ppm and $10$ equally spaced measurements between $z=2.5$ and $z=4.0$, each with an uncertainty $\sigma_\alpha = 0.2$ ppm. These uncertainties are about one order of magnitude better than the statistical uncertainties already demonstrated by the ESPRESSO spectrograph \citep{Murphy}, and should therefore be achievable by ANDES, given the larger collecting area of the telescope, provided the instrument requirements on precision and accuracy are met \citep{HIRES}. These can be seen as the minimal performance of a competitive high-resolution spectrograph in the 2030s, for this science case. For the Optimistic set, we assume $20$ equally spaced measurements between $z=1.0$ and $z=2.5$ each with $\sigma_\alpha =0.05$ ppm $15$ equally spaced measurements between $z=2.5$ and $z=4.0$ each with an uncertainty $\sigma_\alpha =0.1$ ppm, and $15$ equally spaced measurements between $z=4.0$ and $z=7.0$ each with an uncertainty $\sigma_\alpha = 1$ ppm. These assumptions are commensurate with (though not identical to) the assumptions of earlier works \citep{AlvesAlpha,Boas,Euclid}, and they are (as the name implies) somewhat more optimistic than the former in terms of the measurement uncertainties. They are meant to represent datasets that can be obtained using ANDES on the same timescale as its redshift drift measurements. 

Our null cosmological fiducial model is flat $\Lambda$CDM with $h = 0.7$ (where $h$ is the Hubble constant in units of 100 km/s/Mpc), $\Omega_\mathrm{m} = 0.3$, $w_0 = -1$, $w_a = 0$, and vanishing scalar field couplings $\zeta_i$. In the following two sections, we also introduce non-null parametric extensions of this model for each of the two varying $\alpha$ paradigms under study. Similarly to analogous previous studies, we use external priors for each cosmological parameter: $\sigma_h = 0.05$, $\sigma_{\Omega_\mathrm{m}} = 0.03$, $\sigma(w_0)=0.1$ and $\sigma(w_a)=0.4$.

In addition to these cosmological priors we use as prior the most recent result of local laboratory tests with atomic clocks \citep{Lange}, which constrains the local drift rate of $\alpha$, 
\be
\left(\frac{\dot\alpha}{\alpha}\right)_0=(1.0\pm1.1)\times10^{-18}\,\text{yr}^{-1}\,,
\ee
which can alternatively be written, in dimensionless units, as
\begin{equation}
    \frac{1}{H_0}\left(\frac{\dot{\alpha}}{\alpha}\right)_{0} = 0.014 \pm 0.015\, \,\mathrm{ppm}\,,
\end{equation}
one notices that this is a comparatively tight constraint although, as explained above, it provides a second prior on (combinations of) some of the model parameters while not being sensitive to the others. For our purposes the present work, its main importance is that it provides a prior on the scalar field couplings. We note that these models can also be constrained by local Weak Equivalence Principle tests \citep{Bekenstein1,Bekenstein2}. We will not include these tests in the present work, since our goal is to assess the role of ANDES and quantify the improvements in the constraints obtained from the combination of redshift drift and $\alpha$ measurements, as compared to those obtained separately from each of the two probes.

\begin{table*}
    \centering
    \caption{Relevant results of the Fisher Matrix analysis for the Bekenstein model with $\zeta_\mathrm{m}=0$ and $\zeta_\Lambda = 0.1$ ppm. The first set of lines shows the Figures of Merit (rounded to the nearest integer), the following one the correlation coefficients, and the last one the one-sigma marginalized uncertainties. Dri denotes the redshift drift, while Bas and Opt denote the Baseline and Optimistic $\alpha$ datasets described in the text. Couplings are expressed in ppm.}
    \label{tab1}
\begin{tabular}{lcccccc}
\textbf{Parameter}    &  \textbf{Priors}   &  \textbf{Dri} &  \textbf{Bas} &  \textbf{Opt} &  \textbf{Dri+Bas} &  \textbf{Dri+Opt}\\
\toprule
$\boldsymbol{FoM(\Omega_\mathrm{m}, h)}$                    &     290   &     604 &         - &           - &               684 &      685      \\
$\boldsymbol{FoM(\zeta_\mathrm{m},\Omega_\mathrm{m})}$      &       0   &      -  &       766 &        1021 &              1635 &     2162      \\
$\boldsymbol{FoM(\zeta_\Lambda,\Omega_\mathrm{m})}$         &       0   &      -  &        82 &         145 &               176 &      306       \\
$\boldsymbol{FoM(\zeta_\Lambda,\zeta_\mathrm{m})}$          &       0   &      -  &       305 &         777 &               325 &      832       \\
\midrule
$\boldsymbol{\rho(\Omega_\mathrm{m}, h)}$                   &  0.000    &  -0.340 &         - &           - &            -0.305 &   -0.304       \\
$\boldsymbol{\rho(\zeta_\mathrm{m},\Omega_\mathrm{m})}$     &  0.000    &       - &     0.251 &       0.270 &             0.121 &    0.131       \\
$\boldsymbol{\rho(\zeta_\Lambda,\Omega_\mathrm{m})}$        & -0.000    &       - &    -0.107 &      -0.139 &            -0.050 &   -0.066       \\
$\boldsymbol{\rho(\zeta_\Lambda,\zeta_\mathrm{m})}$         & -1.000    &       - &    -0.912 &      -0.930 &            -0.918 &   -0.935       \\
\midrule
$\boldsymbol{\sigma(h)}$                                    &   0.050   &  0.048  &         - &           - &             0.048 &    0.048       \\
$\boldsymbol{\sigma(\Omega_\mathrm{m})}$                    &   0.030   &  0.016  &     0.030 &       0.030 &             0.014 &    0.014       \\
$\boldsymbol{\sigma(\zeta_\mathrm{m})}$                     & 1738.162  &       - &     0.020 &       0.015 &             0.019 &    0.014       \\
$\boldsymbol{\sigma(\zeta_\Lambda)}$                        &22797.643  &       - &     0.178 &       0.102 &             0.177 &    0.102       \\
\end{tabular}
\end{table*}
\begin{figure*}
\begin{center}
\includegraphics[width=1.0\columnwidth,keepaspectratio]{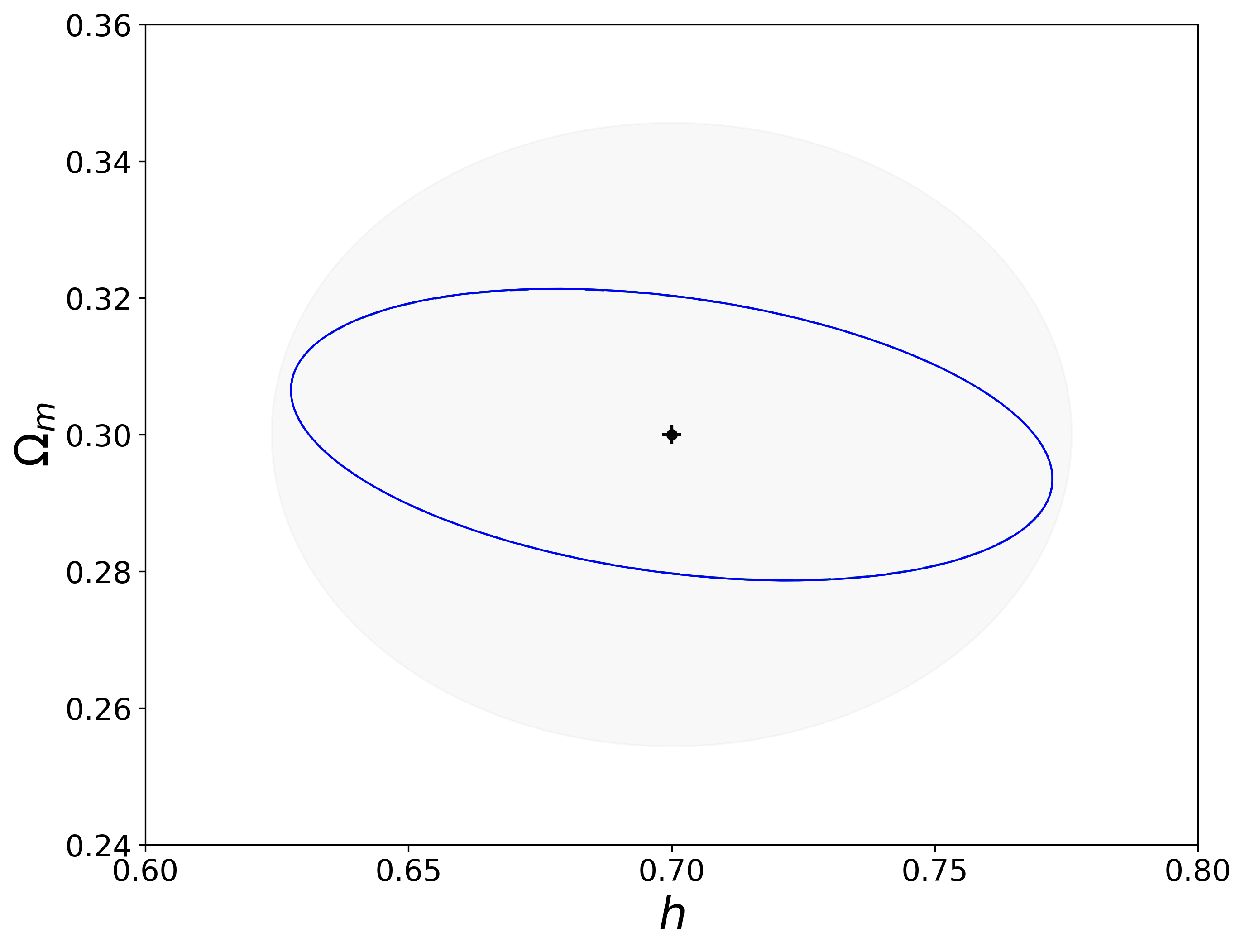}
\includegraphics[width=1.0\columnwidth,keepaspectratio]{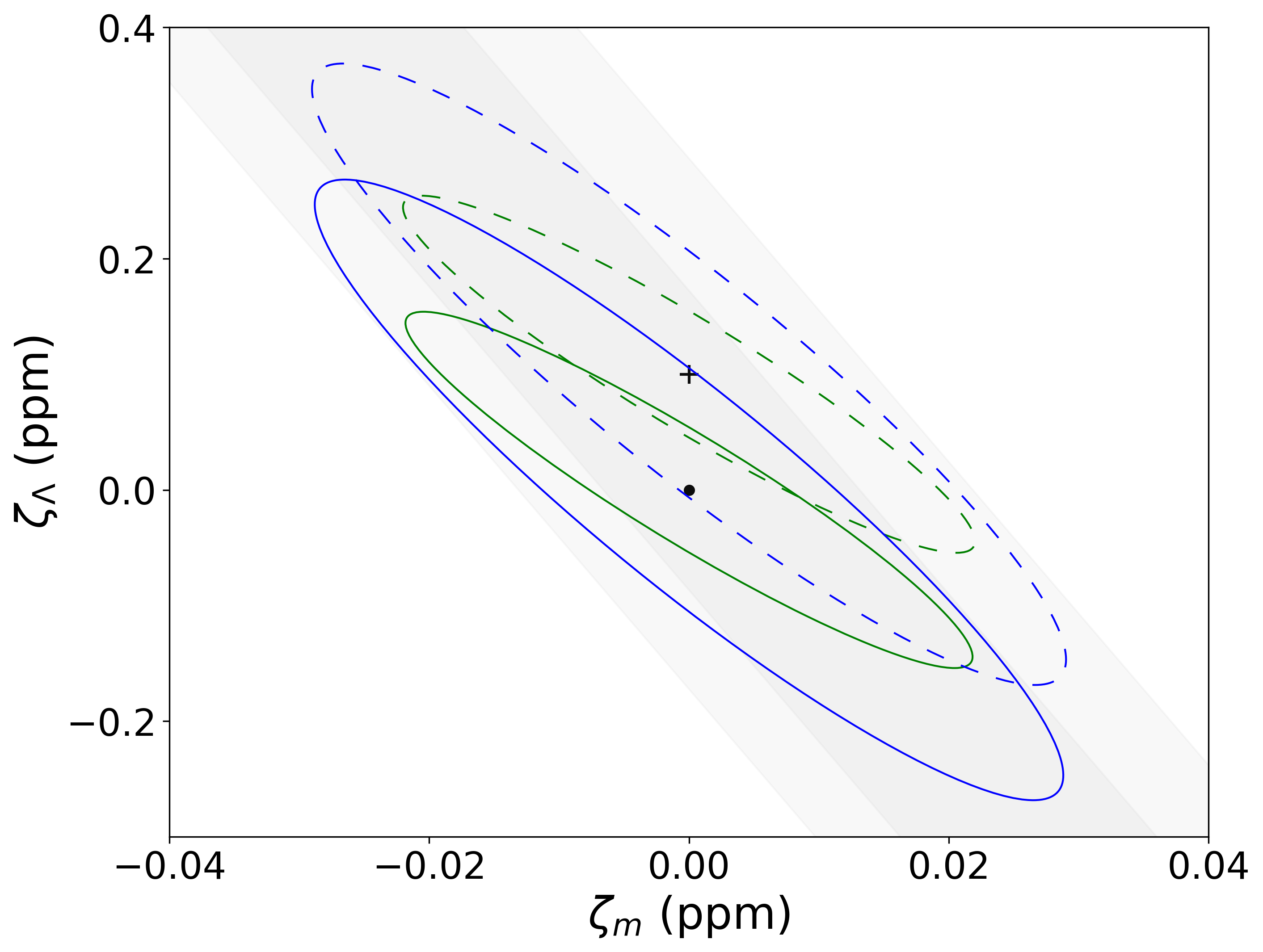}
\end{center}
\caption{Relevant one-sigma constraints from the combination of redshift drift and $\alpha$ measurements for the Bekenstein model. Solid and dashed lines are for the null (standard $\Lambda$CDM) and non-null fiducial models respectively. Baseline and Optimistic assumptions for $\alpha$ measurements are shown in blue and in green respectively. Dots and crosses identify the fiducial values of the parameters and grey areas correspond to the regions allowed by the priors described in the text.}
\label{fig1}
\end{figure*}


\section{Bekenstein-type models}

We start with the Bekenstein models---our fiducial Class II models. This is the simplest class to study since, as we will see, there is little correlation between the cosmological and the fundamental physics parameters (the scalar field couplings). The evolution of $\alpha$ is given by Eq. (\ref{eq:ClasseIImodels}). We consider two specific fiducial models. The first is the null (standard) $\Lambda$CDM, while the second is non-null: it has the same cosmological parameters, but one non-zero coupling, specifically $\zeta_\mathrm{m} = 0$ and $\zeta_{\Lambda}=+0.1$ ppm (parts per million). Both are currently allowed, with the latter saturating existing one-sigma constraints \citep{Bekenstein1,Bekenstein2}.

Table \ref{tab1} shows the main results of the Fisher Matrix analysis for the non-null model: the correlation coefficients for relevant parameter pairs, figures of merit (FoM, defined as the inverse or the area of the corresponding one-sigma confidence ellipse), and one sigma marginalized uncertainties for each model parameter. The table lists the results for each of the two observables as well as for their combination. Given the small correlations between the cosmological parameters and couplings, these values are almost the same for the null fiducial. Figure \ref{fig1} presents the most relevant $2$D plane constraints for the combination of the two probes. For comparison, the table and figure also show the constraints from the priors only; in the latter case these are denoted by the grey areas. Note that in the $\zeta_m$--$\zeta_\Lambda$ plane the priors' Fisher Matrix can be analytically (exactly) calculated and is singular: in other words, its determinant is zero, and therefore it can't be inverted and  formally there are no Fisher matrix constraints on the two individual couplings. Nevertheless this issue (which does not persist when data is added to the matrix) can be circumvented by truncating the numerically calculated the Fisher Matrix to six decimal places. Showing that there are no substantive prior constraints on the individual couplings.

The synergy between the two observables is clearly illustrated. The redshift drift cannot constrain the coupling parameters, being only sensitive to cosmological ones, and especially to the matter density \citep{Alves}: with our assumed dataset, the prior on the Hubble constant is improved by a factor of 1.04 while that of the matter density is improved by a factor of 1.88. Conversely, $\alpha$ data cannot constrain the Hubble constant, and its sensitivity to the matter density, although formally not zero, is negligible. Despite that, it provides separate constraints on both scalar field couplings, while the atomic clocks prior only constrain a combination of the two. The combination of the two datasets allows the model to be fully constrained, in particular improving the FoM in the $\Omega_m$--$\zeta$ planes by more than a factor of two.

We note that the degeneracy direction in the $\zeta_\Lambda$--$\zeta_m$ plane is different for the Baseline and Optimistic $\alpha$ datasets, due to the fact that the sensitivity of $\alpha$ to the two couplings is redshift-dependent \citep{Bekenstein1}. This highlights that one can optimize constraints on models of this kind by a suitable choice of the redshift distribution of the $\alpha$ measurements, when other properties of the targets are equal.

\section{Dynamical dark energy models}

For Class I models we take the CPL parametrization of dark energy. The $\alpha$ variation is given by Eq. (\ref{eq:ClassImodels}), from which we see that, for a $\Lambda$CDM fiducial model, the integral vanishes and there is no cosmological $\alpha$ variation, regardless of the value of $\zeta$. Instead, we consider two non-$\Lambda$CDM fiducials, representative of freezing and thawing canonical scalar field models \citep{Caldwell}; we do not consider the physically problematic phantom fields. Our freezing model has $w_0 = -0.9$ and $w_a = +0.3$ and our thawing model has $w_0 = -0.8$ and $w_a = -0.2$. For both we have $\zeta = -0.15$ ppm, $\Omega_m=0.3$ and $h=0.7$; both models saturate current constraints, and we also use the aforementioned priors. Note that in this case different choices of fiducial models will lead to different correlations between model parameters (and will also be differently affected by the same priors). In particular, as we see in what follows, the results are noticeably different for the freezing and the thawing cases.

\begin{table*}
    \centering
    \caption{Relevant results of the Fisher Matrix analysis for the freezing ($w_0 = -0.9$, $w_a = +0.3$) and thawing ($w_0 = -0.8$, $w_a = -0.2$) CPL models, with $\zeta = -0.15$ ppm. The first set of lines shows the Figures of Merit (rounded to the nearest integer), the following one the correlation coefficients, and the last one the one-sigma marginalized uncertainties. Dri denotes the redshift drift, while Bas and Opt denote the Baseline and Optimistic $\alpha$ datasets described in the text. The coupling is given in ppm.}
    \label{tab2}
\begin{tabular}{ccccccccccccc}
     \multicolumn{6}{|c|}{\textbf{Freezing model}} & \multicolumn{1}{|c|}{\textbf{Parameter} } & \multicolumn{6}{|c|}{\textbf{Thawing model}} \\

\textbf{Dri} &  \textbf{Bas} &  \textbf{Opt} &  \textbf{Dri+Bas} &  \textbf{Dri+Opt} &  \textbf{Priors} 
 &    &  \textbf{Priors}    &  \textbf{Dri} &  \textbf{Bas} &  \textbf{Opt} &  \textbf{Dri+Bas} &  \textbf{Dri+Opt} \\
\toprule
  439  &      - &       - &        589 &       596 &   290   &$\boldsymbol{FoM(\Omega_\mathrm{m}, h)}$       &   290    &     529 &         - &           - &         662 &          673      \\
  220  &    153 &     157 &        432 &       456 &   145   &$\boldsymbol{FoM(w_0, \Omega_\mathrm{m})}$     &   145    &     252 &       149 &         151 &         439 &          449      \\
   69  &     50 &      59 &        138 &       159 &   36    &$\boldsymbol{FoM(w_a, \Omega_\mathrm{m})}$     &   36     &      69 &        62 &          88 &         156 &          218      \\
   13  &     17 &      22 &         32 &        39 &   11    &$\boldsymbol{FoM(w_a, w_0)}$                   &   11     &      12 &        20 &          35 &          34 &           52      \\
    -  &    198 &     209 &        541 &       630 &   177   &$\boldsymbol{FoM(\zeta, \Omega_\mathrm{m})}$   &   329    &       - &       331 &         333 &         865 &          881      \\
    -  &    152 &     162 &        230 &       262 &   132   &$\boldsymbol{FoM(\zeta, w_0)}$                 &   186    &       - &       196 &         201 &         281 &          291      \\
    -  &     23 &      37 &         41 &        63 &   13    &$\boldsymbol{FoM(\zeta, w_a)}$                 &   25     &       - &        43 &          63 &          64 &           89      \\
\midrule
-0.234 &      - &      - &     -0.251 &    -0.258  &   0.000   &$\boldsymbol{\rho(\Omega_\mathrm{m}, h)}$      &   0.000    &  -0.303 &         - &           - &      -0.301 &       -0.309      \\
-0.240 & -0.012 & -0.016 &     -0.258 &    -0.300  &  -0.000   &$\boldsymbol{\rho(w_0, \Omega_\mathrm{m})}$     &   0.000    &  -0.189 &     0.005 &       0.007 &      -0.094 &       -0.074      \\
-0.522 &  0.035 &  0.055 &     -0.352 &    -0.331  &   0.000   &$\boldsymbol{\rho(w_a, \Omega_\mathrm{m})}$    &   0.000    &  -0.378 &     0.030 &       0.052 &      -0.145 &       -0.045      \\
-0.157 &  0.318 &  0.491 &      0.080 &     0.234  &   0.000   &$\boldsymbol{\rho(w_a, w_0)}$                  &   0.000    &  -0.068 &    -0.345 &      -0.616 &      -0.250 &       -0.480      \\
 -     & -0.062 & -0.070 &     -0.320 &    -0.414  &  -0.039   &$\boldsymbol{\rho(\zeta, \Omega_\mathrm{m})}$  &  -0.073    &       - &    -0.071 &      -0.071 &      -0.149 &       -0.152      \\
 -     &  0.911 &  0.909 &      0.835 &     0.832  &   0.916   &$\boldsymbol{\rho(\zeta, w_0)}$                &   0.849    &       - &     0.853 &       0.857 &       0.767 &        0.778      \\
 -     &  0.461 &  0.731 &      0.272 &     0.602  &   0.000   &$\boldsymbol{\rho(\zeta, w_a)}$                &   0.000    &       - &    -0.152 &      -0.278 &      -0.023 &       -0.036      \\
\midrule
0.048  &      - &      - &      0.047 &     0.047  &   0.050   &$\boldsymbol{\sigma(h)}$                       &   0.050    &  0.048  &        -  &           - &        0.048 &        0.048      \\
0.021  &  0.030 &  0.030 &      0.016 &     0.016  &   0.030   &$\boldsymbol{\sigma(\Omega_\mathrm{m})}$       &   0.030    &  0.018  &     0.030 &       0.030 &        0.014 &        0.014      \\
0.096  &  0.095 &  0.092 &      0.065 &     0.063  &   0.100   &$\boldsymbol{\sigma(w_0)}$                     &   0.100    &  0.098  &     0.097 &       0.096 &        0.069 &        0.068      \\
0.349  &  0.292 &  0.245 &      0.210 &     0.182  &   0.400   &$\boldsymbol{\sigma(w_a)}$                     &   0.400    &  0.377  &     0.234 &       0.166 &        0.194 &        0.139      \\
-      &  0.073 &  0.070 &      0.053 &     0.048  &   0.082   &$\boldsymbol{\sigma(\zeta)}$                   &   0.044    &      -  &     0.044 &       0.044 &        0.035 &        0.035      \\
\end{tabular}
\end{table*}
\begin{figure*}
\begin{center}
\includegraphics[width=1.0\columnwidth,keepaspectratio]{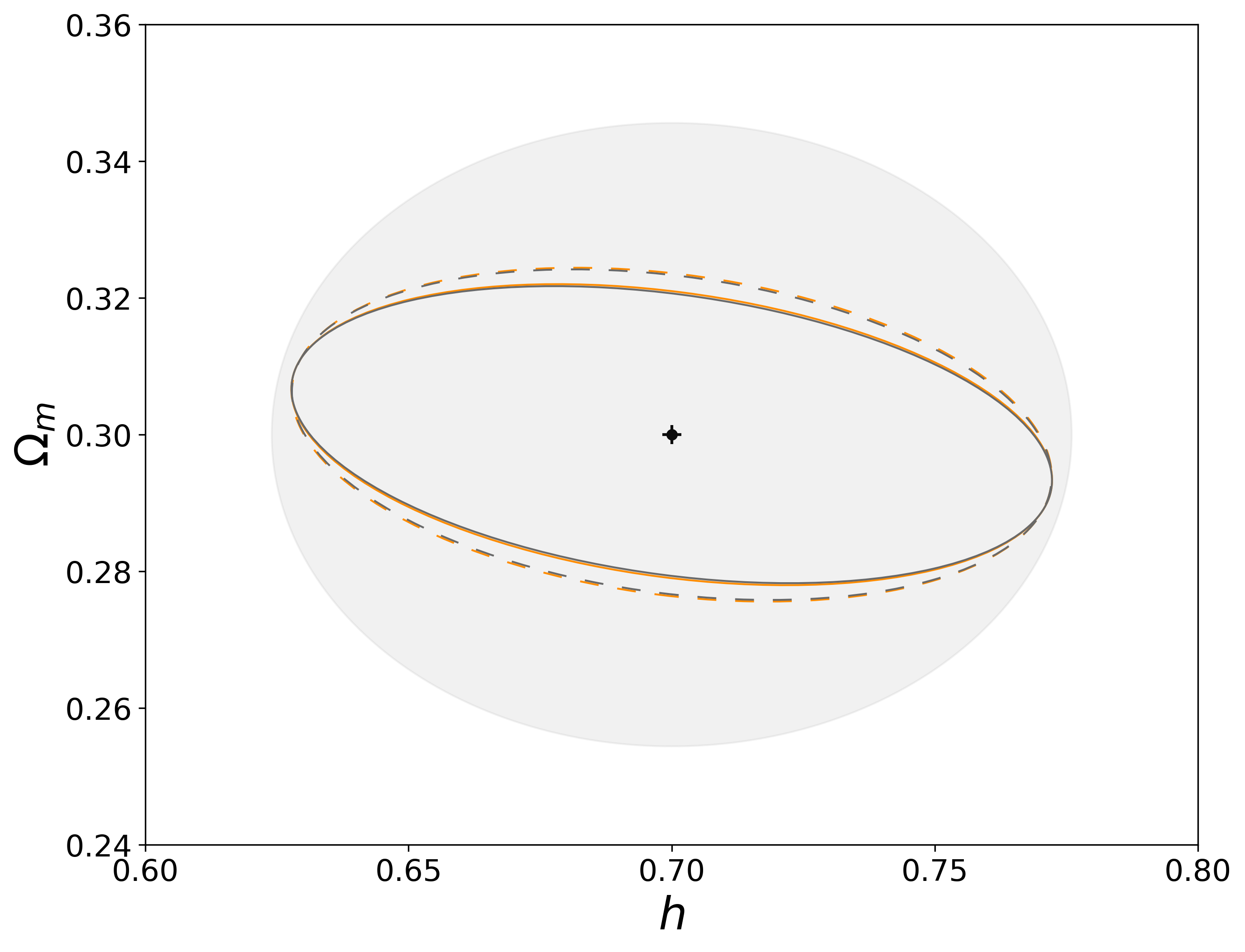}
\includegraphics[width=1.0\columnwidth,keepaspectratio]{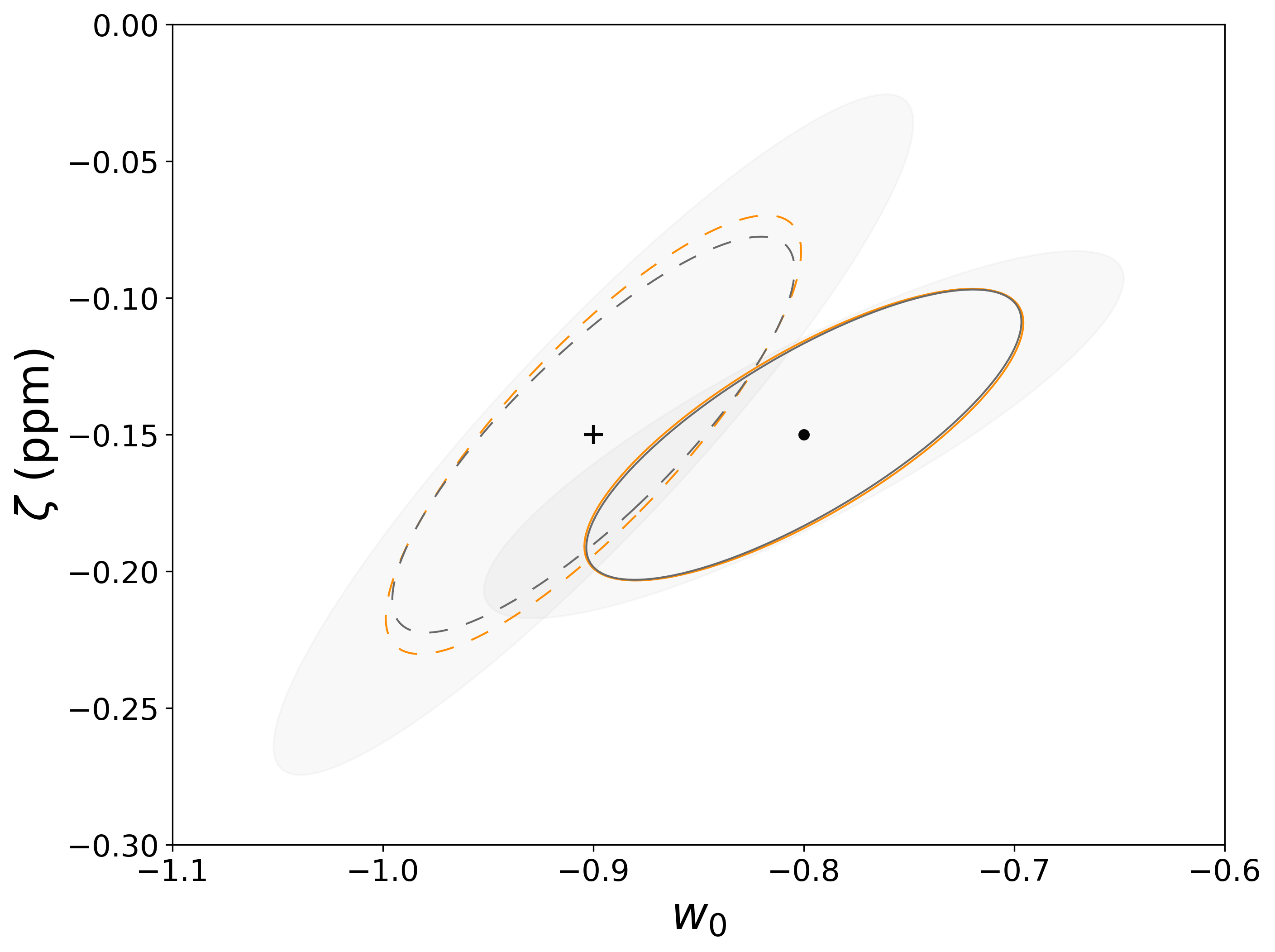}
\includegraphics[width=1.0\columnwidth,keepaspectratio]{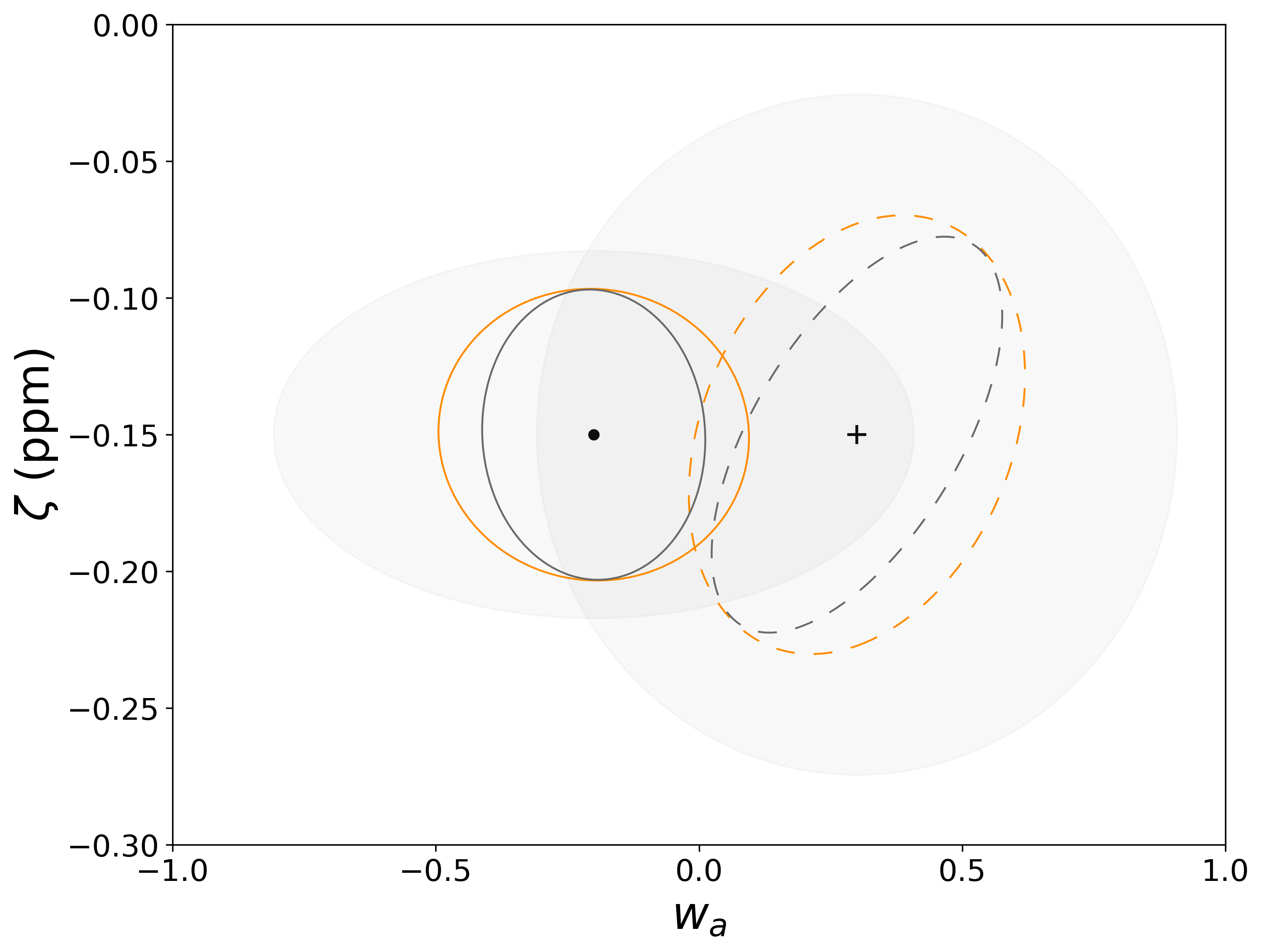}
\end{center}
\caption{One-sigma constraints for the most relevant combinations of the CPL parametrization parameters ($h$, $\Omega_\mathrm{m}$, $w_0$, $w_a$ and $\zeta$). The solid and dashed lines correspond to thawing and freezing models, respectively.  Baseline and Optimistic assumptions for $\alpha$ measurements, combined with redshift drift data, are shown in orange and in black respectively. Dots and crosses identify the fiducial values of the parameters and grey areas correspond to the regions allowed by the priors described in the text.}
\label{fig2}
\end{figure*}

Table \ref{tab2} and Figure \ref{fig2} show the results of the Fisher Matrix analysis for the freezing and thawing cases of this model. Qualitatively the behaviour is similar to that of the Bekenstein case: the redshift drift only constrains the cosmological sector (especially the matter density), while $\alpha$ data constrains the scalar field coupling but also the dark energy equation of state. Regarding this equation of state, $\alpha$ measurements lead to small improvements on the $w_0$ prior (in part because this is already tightly constrained by the atomic clocks prior), but to more significant improvements in $w_a$ (which is not constrained by local data). This also highlights the importance of having measurements across a wide range of redshifts, and thereby mapping the dynamics of a putative scalar filed whose dynamics can be slow. This point is particularly noticeable by comparing the constraints for the Baseline and Optimistic $\alpha$ datasets, bearing in mind that the former span the redshift range from 1 to 4 while the latter span the redshift range from 1 to 7.

FoMs also improve noticeably. For the combined datasets the improvement is always larger than a factor of 2, and in some cases even larger: for the thawing case with the Optimistic $\alpha$ dataset, the figure of merit in the $(w_0,w_a)$ plane improves by more than a factor of 5. Note that correlations between model parameters depend both on the datasets (they are different in the Baseline and Optimistic cases) and on the fiducial model. For example, the dark energy equation of state parameters $w_0$ and $w_a$ are always anti-correlated when redshift drift data is used on its own. For the thawing model the anti-correlation also occurs for $\alpha$ data and for the combination of the two datasets, but for the freezing model the correlation becomes positive whenever $\alpha$ data is present.

It is also worthy of note that in the fundamental physics sector the freezing model constraint on the coupling $\zeta$ is improved by a factor of 1.7 with respect to the prior, while for the thawing model the improvement is only of a factor of 1.3.  This is due to the combination of two separate effects: on the one hand in the freezing model there are larger values of $\alpha$ at non-zero redshift (for the same value of $\zeta$), and on the other hand the atomic clocks prior on $\zeta$ is stronger in the case of the thawing model (by about a factor of 1.9).

\begin{figure*}
\begin{center}
\begin{tabular}{c c}
        \textbf{Freezing model} & \textbf{Thawing model} \\
        \includegraphics[width=1.0\columnwidth,keepaspectratio]{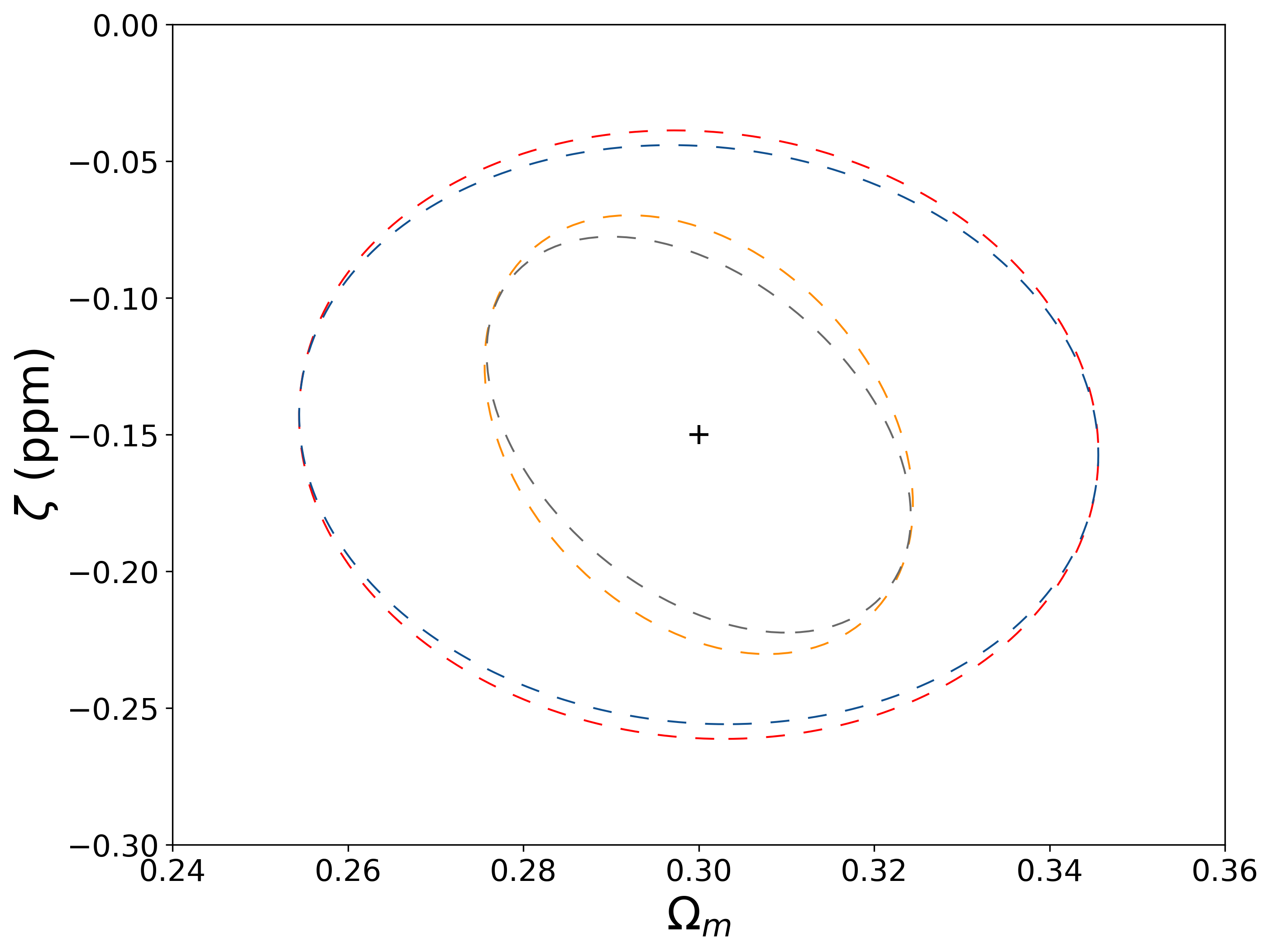} & \includegraphics[width=1.0\columnwidth,keepaspectratio]{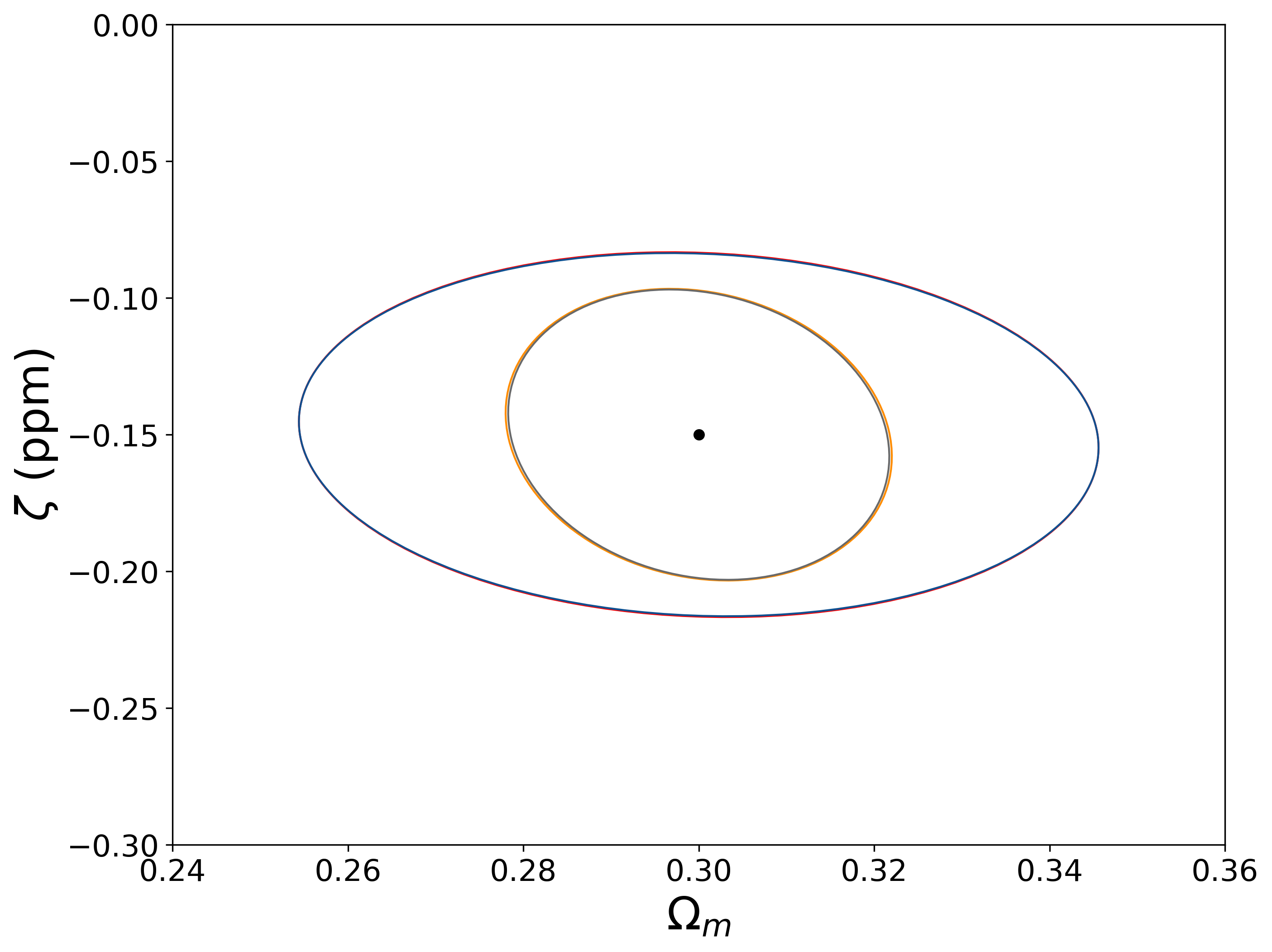}\\
        \includegraphics[width=1.0\columnwidth,keepaspectratio]{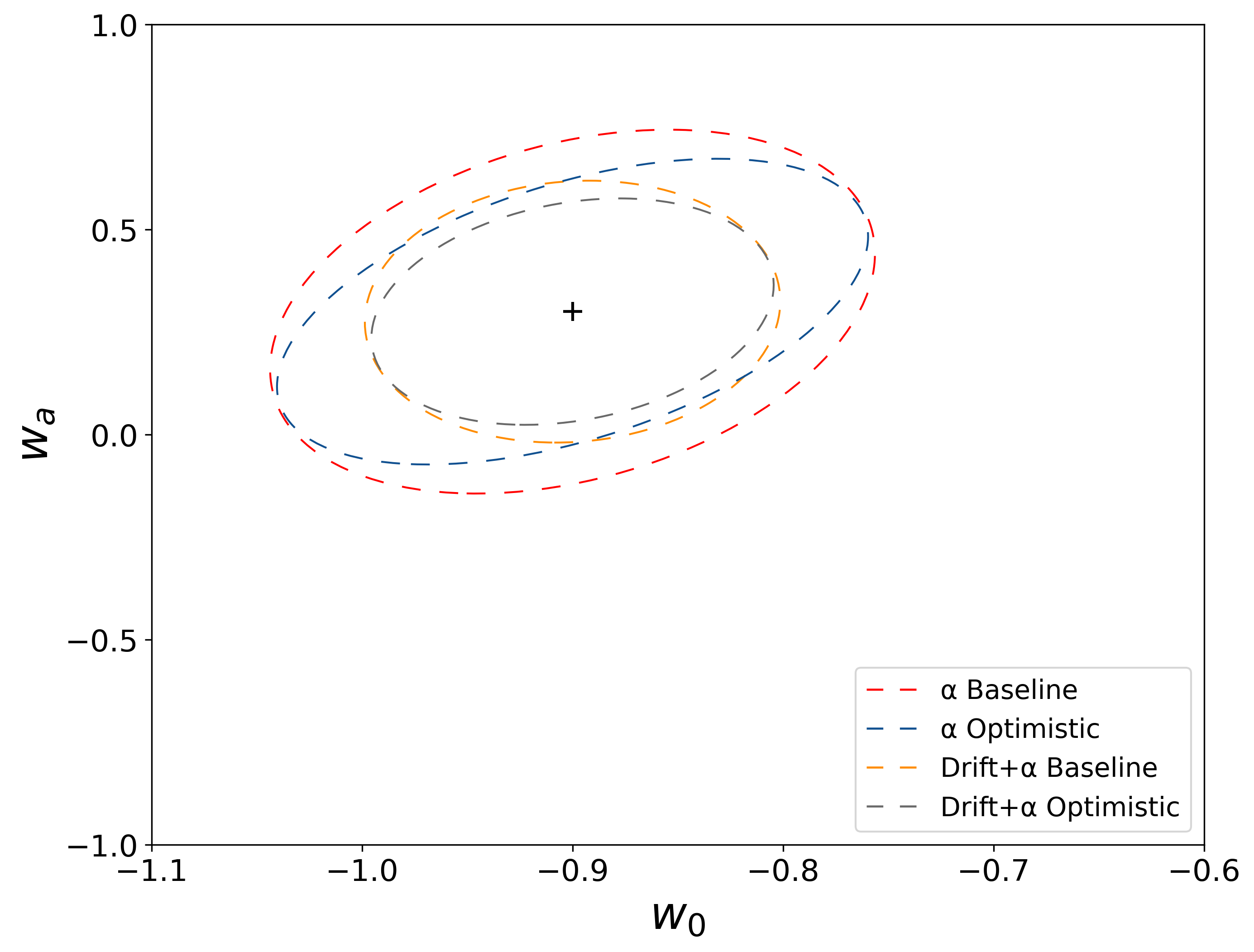} & \includegraphics[width=1.0\columnwidth,keepaspectratio]{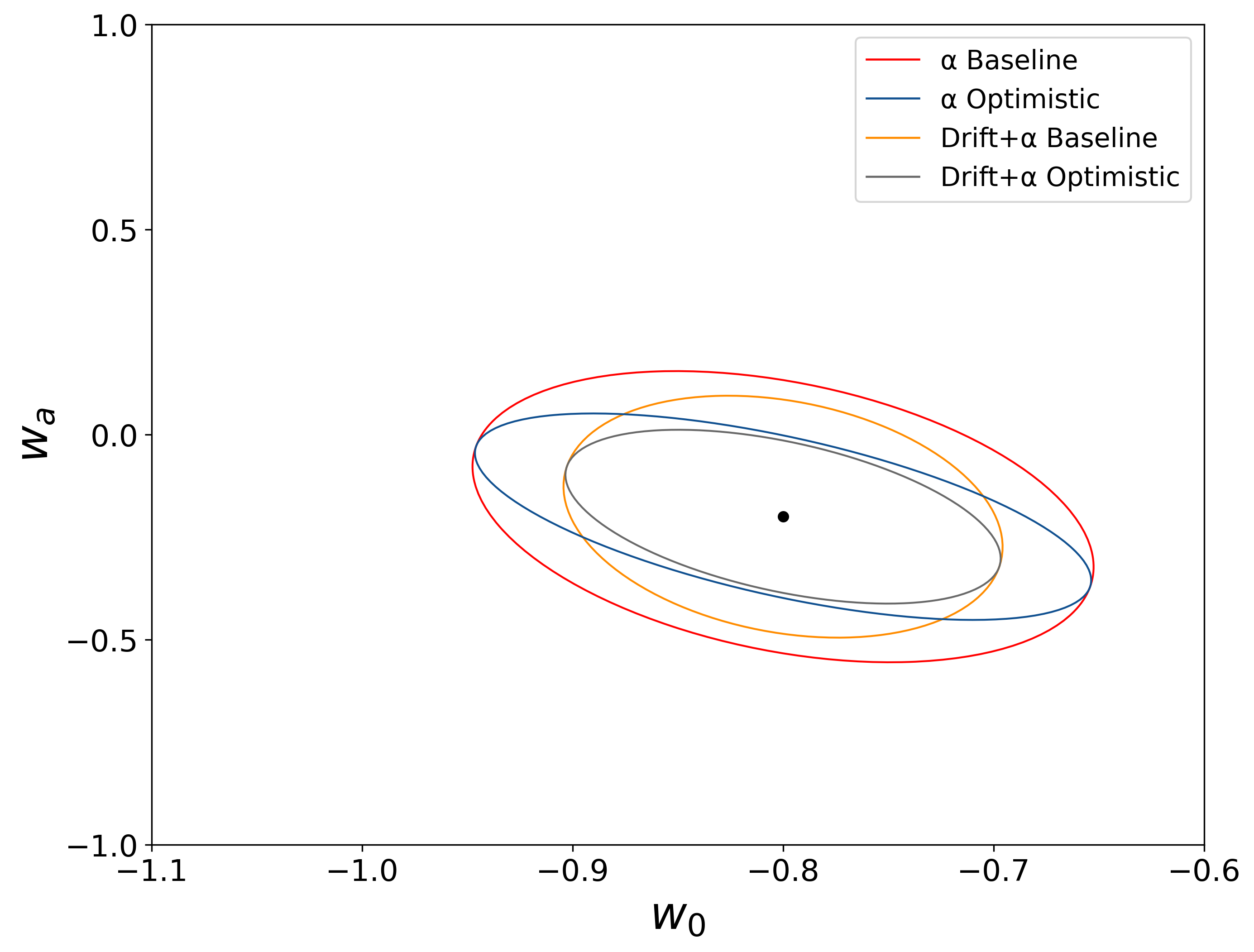}\\
\end{tabular}
\end{center}
\caption{One-sigma constraints of the $\zeta - \Omega_\mathrm{m}$ and $w_0-w_a$ planes for the CPL parametrization. The dashed and solid lines correspond to freezing and thawing models, respectively. Baseline and Optimistic assumptions for $\alpha$ measurements, combined with redshift drift data, are shown in orange and in black respectively. (Both of these choices match those in Fig. \ref{fig2}.) Forecasts for $\alpha$-only measurements are also shown, in red and blue for Baseline and Optimistic assumptions respectively. Dots and crosses identify the fiducial values of the parameters as described in the text. Note that in each pair of panels the axis parameter ranges are the same.}
\label{fig3}
\end{figure*}

Finally, figure \ref{fig3} shows the constraints for the Baseline and Optimistic assumptions for $\alpha$ measurements, alone and in combination with the redshift drift measurement. This highlights both the gains obtained by the addition of the redshift drift measurements, and how these gains depend on the choices of the fiducial model and the $\alpha$ data.

\section{Conclusions}

We have explicitly assessed the synergies between high-resolution astrophysical spectroscopy measurements of the redshift drift and of the fine-structure constant $\alpha$, as will be carried out by the ANDES spectrograph at the ELT. Although the correlations between the cosmological parameters (i.e., the matter density, Hubble constant and dark energy equation of state parameters) and the fundamental physics parameters (i.e., the scalar field couplings) are, in most cases, small, each observable on its own cannot constrain the full model: the redshift drift is not sensitive to the couplings, while $\alpha$ measurements are insensitive to the Hubble constant (and have a modest sensitivity to the matter density). On the other hand, the combination of the two provides constraints on the full set of model parameters.

For the redshift drift, bright high-redshift quasars are scarce, and a key open question relates to the optimal sample size: \citet{Boutsia} proposed a Golden Sample of 30 QSOs, while \citet{Dong} suggests a much smaller one containing only four, and more recently \citet{Cristiani23} proposed a second Golden Sample of seven QSOs. The answer is expected to depend on what is defined as the core scientific goal of the experiment. If the goal is to detect the drift at the highest possible statistical significance, then a small sample will be best. Indeed, were it not for observability constraints, one could have a sample of one: the brightest available target. On the other hand, if the goal is to use the measurements, together with other data, to improve cosmological constraints (and optimally complement those available from other datasets), then being able to make measurements at different redshifts is also an important consideration, since the redshift drift sensitivity to various cosmological parameters is both redshift-dependent and different for different parameters, as previously discussed in \citet{Alves}.

For $\alpha$ (for which a larger number of targets is known) there is an analogous trade-off, between maximizing the chance of a single non-null measurement and constraining the dynamics of putative scalar fields at various redshifts. Both will be studied in detail in the context of the ANDES Phase B. Such detailed optimisation studies are left for subsequent work. Our Fisher Matrix based forecast code, which we publicly release, will be one of the tools for this effort, and in particular it will contribute to the definition of the ANDES observing strategy.

\section*{Acknowledgements}

This work was funded by FEDER---COMPETE 2020 and by FCT through projects POCI-01-0145-FEDER-028987, PTDC/FIS-AST/28987/2017 and 2022.04048.PTDC. CJM also acknowledges FCT and POCH/FSE (EC) support through Investigador FCT Contract 2021.01214.CEECIND/CP1658/CT0001. This work was partially enabled by funding from the UCL Cosmoparticle Initiative.

{\noindent\bf Data availability:} This work uses simulated data, generated as detailed in the text.

\bibliographystyle{mnras}
\bibliography{drift} 

\bsp	
\label{lastpage}
\end{document}